\documentclass[aps,prd,twocolumn,showpacs,groupedaddress]{revtex4} 
\usepackage{graphicx}  % needed for figures
\usepackage{dcolumn}   % needed for some tables
\usepackage{bm}        % for math
\usepackage{amssymb}   % for math

\newcommand{\as}{\alpha_s}
\newcommand{\asmz}{\alpha_s(M_Z)}
\newcommand{\aspT}{\alpha_s(p_T)}
\newcommand{\asmur}{\alpha_s(\mu_r)}
\newcommand{\pythia}{{\sc pythia}}
\newcommand{\fastnlo}{{\sc fastnlo}}
\newcommand{\nlojet}{{\sc nlojet++}}
\newcommand{\ord}{{\cal O}}
\newcommand{\ppbar}{p{\bar{p}}}

\begin{document}
\hspace{5.2in} \mbox{FERMILAB-PUB-09-558-E}

\title{\boldmath  Determination of the strong coupling constant from the 
inclusive jet cross section \\ in $\ppbar$ collisions at $\sqrt{s}=1.96\,$TeV
}
% LIST_OF_AUTHORS_R2.TEX                 10/14/09           
%
\author{V.M.~Abazov$^{37}$}
\author{B.~Abbott$^{75}$}
\author{M.~Abolins$^{65}$}
\author{B.S.~Acharya$^{30}$}
\author{M.~Adams$^{51}$}
\author{T.~Adams$^{49}$}
\author{E.~Aguilo$^{6}$}
\author{M.~Ahsan$^{59}$}
\author{G.D.~Alexeev$^{37}$}
\author{G.~Alkhazov$^{41}$}
% ...

\affiliation{\vspace{0.1 in}(The D\O\ Collaboration)\vspace{0.1 in}}
% ...
\affiliation{$^{6}$University of Alberta, Edmonton, Alberta, Canada;
                Simon Fraser University, Burnaby, British Columbia, Canada;
                York University, Toronto, Ontario, Canada and
                McGill University, Montreal, Quebec, Canada}
% ...
\affiliation{$^{30}$Tata Institute of Fundamental Research, Mumbai, India}
% ...
\affiliation{$^{37}$Joint Institute for Nuclear Research, Dubna, Russia}
% ...
\affiliation{$^{41}$Petersburg Nuclear Physics Institute,
                St. Petersburg, Russia}
% ...
\affiliation{$^{49}$Florida State University, Tallahassee, Florida 32306, USA}
% ...
\affiliation{$^{51}$University of Illinois at Chicago,
                Chicago, Illinois 60607, USA}
% ...
\affiliation{$^{59}$Kansas State University, Manhattan, Kansas 66506, USA}
% ...
\affiliation{$^{65}$Michigan State University,
                East Lansing, Michigan 48824, USA}
% ...
\affiliation{$^{75}$University of Oklahoma, Norman, Oklahoma 73019, USA}
% ...
  % input Dzero author list
\date{Received 13 November 2009; published 29 December, 2009}

\begin{abstract}
We determine the strong coupling constant $\as$ and its energy dependence
from the $p_T$ dependence of the inclusive jet cross section in 
$\ppbar$ collisions at $\sqrt{s}=1.96\,$TeV.
The strong coupling constant is determined over the transverse momentum 
range $50 < p_T < 145\,$GeV.
Using perturbative QCD calculations to order $\ord(\as^3)$
combined with $\ord(\as^4)$ contributions from threshold corrections,
we obtain $\asmz = 0.1161^{+0.0041}_{-0.0048} $.
This is the most precise result obtained at a hadron-hadron collider.
\end{abstract}

\maketitle

% ************************************************************************
% ************   Introduction
% ************************************************************************

Asymptotic freedom, the fact that the strong force between
quarks and gluons keeps getting weaker when it is probed at
increasingly small distances, is a remarkable property 
of quantum chromodynamics (QCD).
This property is reflected by the renormalization group equation (RGE) 
prediction for the dependence of the strong coupling constant $\as$ on 
the renormalization scale $\mu_r$ and therefore on the momentum transfer.
Experimental tests of asymptotic freedom require precise determinations
of $\asmur$ over a large range of momentum transfer.
Frequently, $\as$ has been determined using production rates of hadronic 
jets in either $e^+e^-$ annihilation or in deep-inelastic $ep$ scattering 
(DIS)~\cite{Bethke:2009jm}.
So far there exists only a single $\as$ result from inclusive jet production 
in hadron-hadron collisions.
The CDF Collaboration determined $\as$ from the inclusive jet cross section 
in $\ppbar$ collisions at $\sqrt{s}=1.8\,$TeV obtaining
$\asmz = 0.1178 
        ^{+0.0081}_{-0.0095} (\mbox{exp.}) 
        ^{+0.0071}_{-0.0047} (\mbox{scale}) 
        \pm 0.0059 (\mbox{PDF})    $~\cite{Affolder:2001hn}.

In this article we determine $\as$ and its dependence on the momentum transfer
using the published measurement of the inclusive jet cross 
section~\cite{:2008hua,epaps} with the D0 detector~\cite{d0det} at the 
Fermilab Tevatron Collider in $\ppbar$ collisions at $\sqrt{s}=1.96\,$TeV.
The inclusive jet cross section $d^2\sigma_{\text{jet}}/dp_T d|y|$
was measured using the Run II iterative midpoint cone algorithm~\cite{run2cone}
with a cone radius of 0.7 in rapidity, $y$, and azimuthal angle.
Rapidity is related to the polar scattering angle $\theta$ with respect to 
the beam axis
by $y=0.5 \ln \left[ (1+\beta \cos \theta) / (1-\beta \cos \theta) \right]$
with $\beta=|\vec{p}| / E$.
The measurement comprises 110 data points corrected to the 
particle level~\cite{Buttar:2008jx} and presented as a function of 
the momentum component transverse to the beam direction, $p_T$,
for $p_T>50\,$GeV in six regions of $|y|$ for $0<|y|<2.4$.

% ************************************************************************
% ***************   Theory
% ************************************************************************

The ingredients of perturbative QCD (pQCD) calculations in hadron collisions 
are $\as$, the perturbative coefficients $c_n$ (in the $n$-th power of $\as$),
and the parton distribution functions (PDFs).
Conceptually, PDFs depend only on the hadron momentum fraction $x$ carried 
by the parton and on the factorization scale $\mu_f$.
In practice, PDFs are determined from measurements of observables which 
depend on $\as$.
Therefore resulting PDF parametrizations depend on the assumption for $\as$ 
made in the extraction procedure.
For all precise phenomenology, this implicit $\as$ dependence must be taken 
into account consistently.
The pQCD prediction for the inclusive jet cross section can therefore be 
written as
\begin{equation}
\sigma_{\text{pert}}(\as) = 
     \left(  \sum_n \as^n c_n \right) \otimes f_1(\as) \otimes f_2(\as)
     \, ,
\label{eq:pQCD}
\end{equation}
where the sum runs over all powers $n$ of $\as$ which contribute to the 
calculation ($n=2, 3, 4$ in this analysis, see below).
The $f_{1,2}$ are the PDFs of the initial state hadrons and the 
``$\otimes$'' sign denotes the convolution over the momentum fractions 
$x_1$, $x_2$ of the hadrons. 
Since the RGE uniquely relates the value of $\asmur$ at any scale $\mu_r$
to the value of $\asmz$, all equations can be expressed in terms of $\asmz$. 
The total theory prediction for inclusive jet production is given by the pQCD 
result in (\ref{eq:pQCD}) multiplied by a correction factor for 
nonperturbative effects
\begin{equation}
 \sigma_{\text{theory}}(\asmz) =  
   \sigma_{\text{pert}}(\asmz)
    \cdot  c_{\text{nonpert}}       \, .
\label{eq:allQCD}
\end{equation}
The factor $c_{\text{nonpert}}$ includes corrections due to hadronization 
and the underlying event which have been estimated in Ref.~\cite{:2008hua} 
using \pythia~\cite{pythia} with CTEQ6.5 PDFs~\cite{Tung:2006tb}, 
tune QW~\cite{Albrow:2006rt}, and $\asmz=0.118$.
The hadronization (underlying event) corrections vary between 
-15\% (+30\%) to -3\% (+6\%), 
for $p_T=50$ to $600\,$GeV~\cite{epaps}.

The perturbative results are the sum of a full calculation to $\ord(\as^3)$ 
[next-to-leading order (NLO)], combined with the $\ord(\as^4)$ (2-loop) terms 
from threshold corrections~\cite{Kidonakis:2000gi}.
Adding the 2-loop threshold corrections leads to a significant reduction 
in the $\mu_r$ and $\mu_f$ dependence of the calculation.
The theory calculations are performed in the $\overline{\mbox{MS}}$ 
scheme~\cite{Bardeen:1978yd} for five active quark flavors using the 
next-to-next-to-leading logarithmic (3-loop) approximation of the 
RGE~\cite{Tarasov:1980au,Larin:1993tp}.
The PDFs are taken from the MSTW2008 next-to-next-to-leading order (NNLO) 
parametrizations~\cite{Martin:2009iq,Martin:2009bu} and $\mu_r$ and $\mu_f$ 
are both chosen equal to the jet $p_T$. 
The calculations use \fastnlo~\cite{Kluge:2006xs} based on 
\nlojet~\cite{Nagy:2003tz,Nagy:2001fj} and on code from the authors of 
Ref.~\cite{Kidonakis:2000gi}.

% ************************************************************************
% **********    Concepts
% ************************************************************************

In this analysis, the value of $\as$ is determined from sets of inclusive 
jet cross section data points by minimizing the $\chi^2$ function between 
data and the theory result (\ref{eq:allQCD}) using {\sc minuit}~\cite{minuit}.
Where appropriate, the $\asmz$ result will be evolved to the scale $p_T$ 
using the 3-loop solution of the RGE, providing a result for $\aspT$. 
All correlated experimental and theoretical uncertainties are treated
in the Hessian approach~\cite{Alekhin:2005dx}, except for the $\mu_{r,f}$ 
dependence (see below). 
The central $\asmz$ result is obtained by minimizing $\chi^2$ with respect 
to $\asmz$ and the nuisance parameters for the correlated uncertainties.
By scanning $\chi^2$ as a function of $\asmz$, the uncertainties are 
obtained from the $\asmz$ values for which $\chi^2$ is increased by 1 
with respect to the minimum value.

To determine $\as$ according to this procedure, knowledge of 
$\sigma_{\text{pert}}(\asmz)$ is required as a continuous function of $\asmz$,
over a $\asmz$ range which covers the possible fit results and their 
uncertainties.
This can be achieved based on a series of PDFs obtained under the same 
conditions but for different values of $\asmz$ using interpolation in $\asmz$.
Some recent PDF analyses have applied this strategy and their results 
are documented for different values of $\asmz$.
The MSTW2008 NNLO (NLO) PDF 
parametrizations~\cite{Martin:2009iq,Martin:2009bu} are presented for 
21 $\asmz$ values in the range $0.107 -0.127$ ($0.110 -0.130$) 
in steps of $0.001$ and 
the CTEQ6.6 results~\cite{Nadolsky:2008zw} are available for five values 
of $\asmz = 0.112, 0.114, 0.118, 0.122, 0.125$.
Because of the wide range in $\asmz$ covered by the MSTW2008 PDFs and the fine 
and equidistant spacing in $\asmz$, we use cubic spline interpolation to 
obtain a smooth parametrization for the $\asmz$ dependence of the 
cross section for $0.108 \le \asmz \le 0.126$
($0.111 \le \asmz \le 0.129$) for the NNLO (NLO) PDFs.
This range is sufficient to cover our central values and the uncertainties.
The MSTW2008 analysis includes data sets that have not yet been included in 
other global PDF analyses (DIS jet data from HERA and recent CCFR/NuTeV 
dimuon data);
the results are available in NNLO accuracy which is adequate when including 
the $\ord(\as^4)$ contributions from threshold corrections in the cross 
section calculation.
The CTEQ6.6 PDF parametrizations are available up to NLO, for five $\asmz$ 
values, and for a more limited range in $\asmz$ as compared to MSTW2008.
Therefore the MSTW2008 PDFs are used to obtain the main results for this 
analysis while the CTEQ6.6 PDFs are used for comparison.

% *******************************************************
% **********   Correlations
% *******************************************************

Care must be taken in phenomenological analyses if the observable under 
study was already used to provide significant constraints on the PDFs as
this introduces correlations of experimental and PDF uncertainties, and it 
may affect the sensitivity to possible new physics signals.
Both aspects are relevant in this $\as$ determination since the D0 inclusive 
jet data under study is included in the MSTW2008 PDF analysis.
Since the correlation of experimental and PDF uncertainties is not documented, 
it can not be taken into account when using the PDFs to extract $\asmz$ 
from the jet data.
As a consequence, we must avoid using those jet cross section data points 
which have provided strong PDF constraints.
While the quark PDFs are constrained by precision structure function data,
the only direct source of information on the high $x$ gluon PDF comes 
currently from Tevatron inclusive jet data.
The impact of Tevatron jet data on the gluon density is documented in 
Ref.~\cite{Martin:2009iq} in Figs.~51-53.
Figure 51 shows that excluding the Tevatron jet data starts 
to affect the gluon 
density at $x > 0.2-0.3$, while for $x\lesssim0.25$  the difference in the 
gluon density with and without Tevatron jet data is less than 5\%.
Figure 53 shows that $x<0.3$ is the region in which the gluon results for 
MSTW2008 and CTEQ6.6 are very close.
We conclude that for momentum fractions $x<0.2-0.3$ the Tevatron jet data
do not have a significant impact on the gluon density, and therefore we can 
neglect correlations between PDF and experimental uncertainties for these data.
Based on this constraint we select below those inclusive jet data points 
from which we extract $\as$.

The Tevatron jet data (which access $p_T$ above 500\,GeV) are probing momentum 
transfers at which $\as$ has not yet been probed in other experiments.
Therefore we can not rule out deviations in the running of $\as$ at large 
momentum due to possible new physics contributions to the RGE.
Since such modifications of the RGE are not taken into account in the PDF 
determinations, these effects would effectively be absorbed into the PDFs.
By construction, using such PDFs to extract $\as$ could seemingly confirm 
the RGE expectations, even in the presence of new physics contributions to 
the RGE.
For a consistent $\as$ determination we would therefore exclude high $p_T$ 
data in the region where the RGE has not yet been successfully tested
which is the region of $p_T \gtrsim 200\,$GeV~\cite{Bethke:2009jm}.
However, those data are already removed by the restriction to $x<0.2-0.3$, 
so no additional requirement is needed to account for this.

\begin{figure}[t]
  \centerline{
  \includegraphics[scale=1.06]{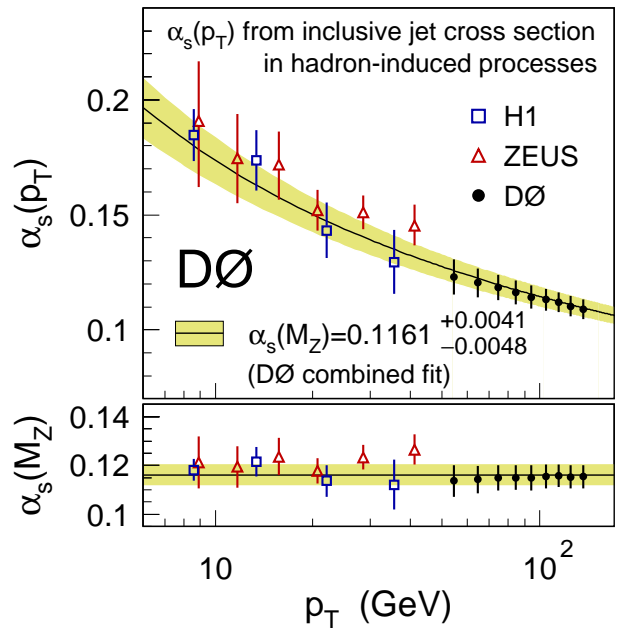}% 
}
\caption{\label{fig:fig1}
  The results for $\aspT$ (top) and $\asmz$ (bottom). 
  The D0 results are based on 22 selected data points which
  have been grouped to produce the 9 data points shown.
  For comparison, results from HERA DIS jet data have been included
  and also the RGE prediction for the combined D0 fit result and 
  its uncertainty (line and band).
  All data points are shown with their total uncertainties.
}\end{figure}

\begin{table*}
\caption{\label{tab:results}
Central values and uncertainties due to different sources
for the nine $\aspT$ results and for the 
combined $\asmz$ result (bottom).
All uncertainties are multiplied by a factor of $10^3$.}
\begin{ruledtabular}
\begin{tabular}{lccccccccc}
$p_T$ range  & No.~of & 
$p_T$  &   &
Total  &
Experimental  &   Experimental  &    Nonperturb.  & PDF   & $\mu_{r,f}$ \\
(GeV)   & data points  &  (GeV)  &  $\aspT$ & uncertainty & uncorrelated & correlated &
correction & uncertainty & variation \\
\hline
50--60  & 4 &  \phantom{1}54.5    &   0.1229  & $^{+7.6}_{-7.7}$ 
& $\pm0.4$
& $^{+4.8}_{-4.9}$ 
& $^{+5.8}_{-5.6}$    
& $^{+0.4}_{-0.6}$ 
& $^{+1.0}_{-1.9}$ 
\\
60--70  & 4 &   \phantom{1}64.5   &   0.1204    & $^{+6.2}_{-6.3}$ 
& $\pm0.3$
& $^{+4.1}_{-4.3}$ 
& $^{+4.5}_{-4.3}$ 
& $^{+0.6}_{-0.5}$ 
& $^{+1.3}_{-1.5}$ 
\\
70--80 & 3 & \phantom{1}74.5  &   0.1184    & $^{+5.6}_{-5.6}$ 
& $\pm0.3$
& $^{+3.8}_{-3.9}$ 
& $^{+4.0}_{-3.9}$
& $^{+0.6}_{-0.6}$ 
& $^{+1.0}_{-0.9}$ 
\\
80--90  & 3 & \phantom{1}84.5   &  0.1163     & $^{+5.1}_{-5.1}$ 
& $\pm0.3$
& $^{+3.6}_{-3.7}$ 
& $^{+3.5}_{-3.5}$
& $^{+0.7}_{-0.7}$ 
& $^{+0.9}_{-0.6}$ 
\\
90--100  & 2 & \phantom{1}94.5   &   0.1142    & $^{+5.1}_{-4.9}$ 
& $\pm0.3$
& $^{+3.5}_{-3.6}$ 
& $^{+3.5}_{-3.3}$
& $^{+0.8}_{-0.8}$ 
& $^{+1.1}_{-0.6}$ 
\\
100--110 &  2 & 104.5  &  0.1131    & $^{+4.7}_{-4.7}$ 
& $\pm0.2$
& $^{+3.4}_{-3.5}$ 
& $^{+3.1}_{-3.0}$
& $^{+0.8}_{-0.8}$ 
& $^{+1.1}_{-0.6}$ 
\\
110--120 &  2 &  114.5  &   0.1121    & $^{+4.2}_{-4.4}$ 
& $\pm0.2$
& $^{+3.1}_{-3.3}$ 
& $^{+2.5}_{-2.7}$
& $^{+0.7}_{-0.8}$ 
& $^{+1.2}_{-0.7}$ 
\\
120--130 &  1 &  124.5  &   0.1102    & $^{+4.4}_{-4.4}$ 
& $\pm0.2$
& $^{+3.2}_{-3.4}$ 
& $^{+2.6}_{-2.6}$
& $^{+0.9}_{-0.9}$
& $^{+1.4}_{-0.9}$  
\\ 
130--145 & 1 &   136.5  &   0.1090    & $^{+4.2}_{-4.3}$ 
& $\pm0.3$
& $^{+3.1}_{-3.4}$ 
& $^{+2.3}_{-2.4}$ 
& $^{+0.9}_{-0.9}$ 
& $^{+1.5}_{-0.9}$
\\ 
% \hline
50--145    & 22
 &   $M_Z$  & 0.1161    & $^{+4.1}_{-4.8}$  
& $\pm0.1$
& $^{+3.4}_{-3.3}$ 
& $^{+1.0}_{-1.6}$ 
& $^{+1.1}_{-1.2}$ 
& $^{+2.5}_{-2.9}$ 
\\
\end{tabular}
\end{ruledtabular}
\end{table*}

% ******************************************************************
% **********     Defining the data set
% ******************************************************************

In $2 \rightarrow 2$ processes, given the rapidities and $p_T$ of the 
two jets, one can compute the momentum fractions $x_1$ and $x_2$ carried 
by the initial partons.
The inclusive jet cross section at given $p_T$ and $|y|$ is, however, 
integrated over all additional jets in an event, so the rapidity of the other 
jet and therefore the full event kinematics, including $x_1$ and $x_2$, 
are not known.
The value of the larger momentum fraction $x_{\text max} = \max(x_1,x_2)$ can 
be computed only under an assumption for the rapidity of the unobserved jet.
For each inclusive jet ($p_T, |y|$) bin we define the variable 
$\tilde{x} = x_T \cdot (e^{|y|}+1)/2$ where $x_T = 2p_T/\sqrt{s}$,
$p_T$ is taken at the bin center, and $|y|$ at the lower boundary of 
the $|y|$ bin.
This variable $\tilde{x}$ corresponds to $x_{\text max}$ for the case that 
the unobserved jet was produced at $y=0$.
In the pQCD calculation, for a given inclusive jet ($p_T, |y|$) bin the 
distribution of $x_{\text max} = \max(x_1,x_2)$ always has a peak plus a 
tail towards high $x_{\text max}$ values.
Although the variable $\tilde{x}$ does not represent the peak position of 
the $x_{\text max}$ distribution, it is correlated with that distribution.
The requirement $\tilde{x} < 0.15$ removes all data points for which more 
than half of the cross section is produced at $x_{\text max} \gtrsim 0.25$.
This leaves 22 (out of 110) data points for the $\as$ analysis with
$p_T < 145\,$GeV for $0<|y|<0.4$,
$p_T < 120\,$GeV for $0.4<|y|<0.8$,
$p_T < 90\,$GeV for $0.8<|y|<1.2$, and
$p_T < 70\,$GeV for $1.2<|y|<1.6$.
Although this selection criterion is well-motivated, the specific choices of
the variable $\tilde{x}$ and the requirement $\tilde{x}< 0.15$ are somewhat 
arbitrary.
We have therefore studied variations of the selection requirement in the 
range $\tilde{x} < 0.10-0.17$ and other choices for the definition of 
$\tilde{x}$ (for example assuming that the unobserved jet has $y_2 = \pm |y|$),
and, we find that the $\as$ results are stable within 1\%.
We conclude that the choice of $\tilde{x}<0.15$ restricts the jet data to 
those points which receive no significant contributions from 
$x_{\text{max}} > 0.25$.
For these data points, experimental and PDF uncertainties are treated as 
being uncorrelated.

% ******************************************************************
% **********    Uncertainties
% ******************************************************************

In the $\as$ determination, we consider the uncorrelated experimental 
uncertainties and all 23 sources of correlated experimental uncertainties
as documented in Refs.~\cite{:2008hua,epaps}.
The nonperturbative corrections are divided into hadronization and 
underlying event effects.
The uncertainty for each is taken to be half the size of the corresponding 
effect.
PDF uncertainties are computed using the 20 68\%~C.L.~uncertainty 
eigenvectors as provided by MSTW2008~\cite{Martin:2009iq}.
The uncertainties in the pQCD calculation due to uncalculated higher order 
contributions are estimated from the $\mu_{r,f}$ dependence of the 
calculations when varying the scales in the range $0.5 < \mu_{r,f}/p_T < 2$.
In the kinematic region under study, variations of $\mu_r$ and $\mu_f$ 
have positively correlated effects on the jet cross sections.
A correlated variation of both scales is therefore a conservative estimate 
of the corresponding uncertainty.
Since the $\mu_{r,f}$ uncertainties can not be treated as Gaussian, these 
are not included in the Hessian $\chi^2$ definition.
Following Refs.~\cite{:2007pb,Chekanov:2006yc}, the $\as$ fits are repeated 
for different choices ($\mu_{r,f} = 0.5p_T$ and $\mu_{r,f} = 2p_T$)
and the differences to the central result (obtained for $\mu_{r,f} = p_T$) 
are taken to be the corresponding uncertainties for $\asmz$. 
Those are added in quadrature to the other uncertainties to obtain the 
total uncertainty.

% *******************************************************
% *********    Results
% *******************************************************

Data points from different $|y|$ regions with similar $p_T$ are grouped to 
determine the results for $\asmz$ and $\aspT$.
A combined fit to all 22 data points yields 
$\asmz =  0.1161^{+0.0041}_{-0.0048}$ with $\chi^2 / \mbox{Ndf} = 17.2/21$.
The results are shown in Fig.~\ref{fig:fig1} as nine $\aspT$ (top) and 
$\asmz$ values (bottom) in the range $50<p_T<145\,$GeV with their total 
uncertainties which are largely correlated between the points.
Also included are results at lower $p_T$ from inclusive jet cross sections 
in DIS from the HERA experiments H1~\cite{:2007pb} and 
ZEUS~\cite{Chekanov:2006yc} and the 3-loop RGE prediction for 
our combined $\asmz$ result. 
Our $\aspT$ results are consistent with the energy dependence predicted 
by the RGE and extend the HERA results towards higher $p_T$.
The combined result is consistent with the result of 
$\asmz = 0.1189 \pm 0.0032$ from combined HERA jet data~\cite{Glasman:2007sm} 
and with the world average value of 
$\asmz = 0.1184\pm0.0007$~\cite{Bethke:2009jm}.
The contributions from individual uncertainty sources are listed in 
Table~\ref{tab:results}. 
The largest source is the experimental correlated uncertainty for which 
the dominant contributions are from the jet energy calibration, the $p_T$ 
resolution and the integrated luminosity.

Varying the size of the uncertainties of the nonperturbative corrections
between a factor of $0.5$ and $2$ changes the central value by 
$^{+0.0003}_{-0.0010}$
and does not affect the uncertainty of the combined $\asmz$ result.
Replacing the MSTW2008 NNLO PDFs by the CTEQ6.6 PDFs changes the central 
result by only $+0.5\%$ which is much less than the PDF uncertainty.
Excluding the 2-loop contributions from threshold corrections and using 
pure NLO pQCD (together with MSTW2008 NLO PDFs and the 2-loop RGE)
gives a result of $\asmz = 0.1202^{+0.0072}_{-0.0059}$.
The small increase in the central value is a result of the missing 
$\ord(\as^4)$ contributions which are compensated by a corresponding increase 
in $\as$.
The difference to the central result is well within the scale uncertainty of 
the NLO result.
The increased uncertainty is mainly caused by the increased $\mu_{r,f}$ 
dependence, but also by the larger PDF uncertainty at NLO.

% ***********************************************************
% **********    Summary
% ***********************************************************

In summary, we have determined the strong coupling constant from the 
inclusive jet cross section using theory prediction in NLO plus 
2-loop threshold corrections.
The $\aspT$ results support the energy dependence predicted by the 
renormalization group equation.
The combined result from 22 selected data points is
$\asmz =  0.1161^{+0.0041}_{-0.0048}$.
This is the most precise $\as$ result obtained at a hadron collider.

% acknowledgement_paragraph_r2.tex                         10/14/09
%
We thank Graeme Watt for helpful discussions.
We thank the staffs at Fermilab and collaborating institutions, 
and acknowledge support from the 
DOE and NSF (USA);
CEA and CNRS/IN2P3 (France);
FASI, Rosatom and RFBR (Russia);
CNPq, FAPERJ, FAPESP and FUNDUNESP (Brazil);
DAE and DST (India);
Colciencias (Colombia);
CONACyT (Mexico);
KRF and KOSEF (Korea);
CONICET and UBACyT (Argentina);
FOM (The Netherlands);
STFC and the Royal Society (United Kingdom);
MSMT and GACR (Czech Republic);
CRC Program, CFI, NSERC and WestGrid Project (Canada);
BMBF and DFG (Germany);
SFI (Ireland);
The Swedish Research Council (Sweden);
and
CAS and CNSF (China).


\begin{thebibliography}{99}
  % list_of_visitor_addresses_r2.tex                         10/14/09
%  available symbols are:
%  $\ast, \dag, \ddag, \S, \P, $\|$, $\ast\ast$, \dag\dag, \ddag\ddag ,\#
%
\bibitem[a]{alton}
Visitor from Augustana College, Sioux Falls, SD, USA.
\bibitem[b]{atramentov,gershtein}
Visitor from Rutgers University, Piscataway, NJ, USA.
\bibitem[c]{burdin}
Visitor from The University of Liverpool, Liverpool, UK.
\bibitem[d]{haas}
Visitor from SLAC, Menlo Park, CA, USA.
\bibitem[e]{luna-garcia}
Visitor from Centro de Investigacion en Computacion - IPN,
  Mexico City, Mexico.
\bibitem[f]{podesta-lerma}
Visitor from ECFM, Universidad Autonoma de Sinaloa, Culiac\'an, Mexico.
\bibitem[g]{weber}
Visitor from Universit{\"a}t Bern, Bern, Switzerland.
\bibitem[h]{wenger}
Visitor from Universit{\"a}t Z{\"u}rich, Z{\"u}rich, Switzerland.
%\bibitem[?]{coadou}
%Visitor from CERN, Geneva, Switzerland.
%\bibitem[?]{hooper}
%Visitor from Bradley University, Peoria, IL, USA.
%\bibitem[?]{kozminski}
%Visitor from Lewis University, Romeoville, IL, USA.
%\bibitem[\ddag]{deceased}
%Deceased.

%
\vskip 0.25cm
  % input visitors address


%\cite{Bethke:2009jm}
\bibitem{Bethke:2009jm}
  S.~Bethke,
  %``The 2009 Wolrd Average of $\alpha_s (M_Z)$,''
  Eur.\ Phys.\ J.\  C {\bf 64}, 689 (2009).
%  [arXiv:0908.1135 [hep-ph]].
  %%CITATION = EPHJA,C64,689;%%

%\cite{Affolder:2001hn}
\bibitem{Affolder:2001hn}
   T.~Affolder {\it et al.} (CDF Collaboration),
  %``Measurement of the strong coupling constant from inclusive jet production
  %at the Tevatron $\bar{p}p$ collider,''
  Phys.\ Rev.\ Lett.\  {\bf 88}, 042001 (2002).
%  [arXiv:hep-ex/0108034].
  %%CITATION = PRLTA,88,042001;%%


%\cite{:2008hua}
\bibitem{:2008hua}
  V.~M.~Abazov {\it et al.} (D0 Collaboration),
 %``Measurement of the inclusive jet cross-section in $p \bar{p}$ collisions at
  %$s^{91/2)}$ =1.96-TeV,''
  Phys.\ Rev.\ Lett.\  {\bf 101}, 062001 (2008).
%  [arXiv:0802.2400 [hep-ex]].
  %%CITATION = PRLTA,101,062001;%%


\bibitem{epaps}
See EPAPS Document No. E-PRLTAO-101-033833 for tables of 
the inclusive jet cross section results and the 
uncertainties. 
For more information on EPAPS see
http://www.aip.org/pubservs/epaps.html


\bibitem{d0det}  %  --- Standard D0 detector reference
   V.~M.~Abazov {\it et al.} (D0 Collaboration), 
  Nucl. Instrum. Methods Phys. Res. A {\bf 565}, 463  (2006).


\bibitem{run2cone} G.~C.~Blazey {\it et al.}, in
     {\sl Proceedings of the Workshop:
     ``QCD and Weak Boson Physics in Run II''},
     Batavia, Illinois, 2000,
     edited by U.~Baur, R.~K.~Ellis, and D.~Zeppenfeld, 
     (FERMILAB Report No.\ FERMILAB-PUB-00-297),
     p 47, see Section 3.5.


\bibitem{Buttar:2008jx}       % particle level
  C.~Buttar {\it et al.},
  arXiv:0803.0678,  Sec. 9.
  %%CITATION = ARXIV:0803.0678;%%


\bibitem{pythia} T.~Sj\"ostrand  {\it et al.},
               Comput.~Phys.~Commun.~135, 238 (2001).          


%\cite{Tung:2006tb}            % CTEQ6.5
\bibitem{Tung:2006tb}
  W.~K.~Tung {\it et al.}  
      %, H.~L.~Lai, A.~Belyaev, J.~Pumplin, D.~Stump and C.~P.~Yuan,
  %``Heavy quark mass effects in deep inelastic scattering and global QCD
  %analysis,''
  JHEP {\bf 0702}, 053 (2007).
%  [arXiv:hep-ph/0611254].
  %%CITATION = JHEPA,0702,053;%%


%\cite{Albrow:2006rt}                      % tune QW
\bibitem{Albrow:2006rt}
  M.~G.~Albrow {\it et al.}  (TeV4LHC QCD Working Group),
  %``Tevatron-for-LHC Report of the QCD Working Group,''
  FERMILAB Report No.\ FERMILAB-CONF-06-359,
  arXiv:hep-ph/0610012.
  %%CITATION = HEP-PH/0610012;%%


%\cite{Kidonakis:2000gi}
\bibitem{Kidonakis:2000gi}
  N.~Kidonakis and J.~F.~Owens,
  %``Effects of higher-order threshold corrections in high-E(T) jet
  %production,''
  Phys.\ Rev.\  D {\bf 63}, 054019 (2001).
%  [arXiv:hep-ph/0007268].
  %%CITATION = PHRVA,D63,054019;%%


%\cite{Bardeen:1978yd}       % MS-bar
\bibitem{Bardeen:1978yd}
  W.~A.~Bardeen, A.~J.~Buras, D.~W.~Duke and T.~Muta,
  %``Deep Inelastic Scattering Beyond The Leading Order In Asymptotically Free
  %Gauge Theories,''
  Phys.\ Rev.\  D {\bf 18}, 3998 (1978).
  %%CITATION = PHRVA,D18,3998;%%


%\cite{Tarasov:1980au}
\bibitem{Tarasov:1980au}
  O.~V.~Tarasov, A.~A.~Vladimirov and A.~Y.~Zharkov,
  %``The Gell-Mann-Low Function Of QCD In The Three Loop Approximation,''
  Phys.\ Lett.\  B {\bf 93}, 429 (1980).
  %%CITATION = PHLTA,B93,429;%%


%\cite{Larin:1993tp}
\bibitem{Larin:1993tp}
  S.~A.~Larin and J.~A.~M.~Vermaseren,
  %``The Three Loop QCD Beta Function And Anomalous Dimensions,''
  Phys.\ Lett.\  B {\bf 303}, 334 (1993).
%  [arXiv:hep-ph/9302208].
  %%CITATION = PHLTA,B303,334;%%


%\cite{Martin:2009iq}
\bibitem{Martin:2009iq}           % --- MSTW2008
  A.~D.~Martin, W.~J.~Stirling, R.~S.~Thorne and G.~Watt,
  %``Parton distributions for the LHC,''
  Eur.\ Phys.\ J.\  C {\bf 63}, 189 (2009).
  %[arXiv:0901.0002 [hep-ph]].
  %%CITATION = EPHJA,C63,189;%%


%\cite{Martin:2009bu}
\bibitem{Martin:2009bu}
  A.~D.~Martin, W.~J.~Stirling, R.~S.~Thorne and G.~Watt,
  %``Uncertainties on alpha_S in global PDF analyses,''
  Eur.\ Phys.\ J.\  C {\bf 64}, 653 (2009).
%  [arXiv:0905.3531 [hep-ph]].
  %%CITATION = EPHJA,C64,653;%%


%\cite{Kluge:2006xs}
\bibitem{Kluge:2006xs}
  T.~Kluge, K.~Rabbertz and M.~Wobisch,
  %``fastNLO: Fast pQCD calculations for PDF fits,''
  DESY Report No.\ DESY-06-186, FERMILAB Report No.\ FERMILAB-CONF-06-352-E,
  arXiv:hep-ph/0609285.
  %%CITATION = HEP-PH 0609285;%%


%%\cite{Nagy:2003tz}
\bibitem{Nagy:2003tz}
  Z.~Nagy,
%   ``Next-to-leading order calculation of three-jet observables in hadron
  %hadron collision,''
  Phys.\ Rev.\ D {\bf 68}, 094002 (2003).
%  [arXiv:hep-ph/0307268].
  %%CITATION = HEP-PH 0307268;%%


%\cite{Nagy:2001fj}
\bibitem{Nagy:2001fj}
  Z.~Nagy,
 %  ``Three-jet cross sections in hadron hadron collisions at next-to-leading
  %order,''
  Phys.\ Rev.\ Lett.\  {\bf 88}, 122003 (2002).
%  [arXiv:hep-ph/0110315].
  %%CITATION = HEP-PH 0110315;%%



\bibitem{minuit} 
F.~James, 
%"Minuit, Function Minimization and Error Analysis", 
CERN long writeup D506.


%\bibitem{hessian}  Hessian $\chi^2$ definition from dijet chi paper
\bibitem{Alekhin:2005dx}
 A.~Cooper-Sarkar and C.~Gwenlan,
in {\sl Proceedings of the Workshop: HERA and the LHC, Part A,
Geneva, Switzerland, 2005},
edited by A. De Roeck and H. Jung, 
(CERN Report No.\ CERN-2005-014, DESY Report No.\ DESY-PROC-2005-01)
see Part 2, Sec. 3.
%  arXiv:hep-ph/0601012.


%\cite{Nadolsky:2008zw}           %  CTEQ6.6
\bibitem{Nadolsky:2008zw}
  P.~M.~Nadolsky {\it et al.},
  %``Implications of CTEQ global analysis for collider observables,''
  Phys.\ Rev.\  D {\bf 78}, 013004 (2008).
%  [arXiv:0802.0007 [hep-ph]].
  %%CITATION = PHRVA,D78,013004;%%


%\cite{:2007pb}
\bibitem{:2007pb}
  A.~Aktas {\it et al.} (H1 Collaboration), 
  %``Measurement of Inclusive Jet Production in Deep-Inelastic Scattering at
  %High Q^2 and Determination of the Strong Coupling,''
  Phys.\ Lett.\  B {\bf 653}, 134 (2007).
%  [arXiv:0706.3722 [hep-ex]].
  %%CITATION = PHLTA,B653,134;%%


%\cite{Chekanov:2006yc}
\bibitem{Chekanov:2006yc}
  S.~Chekanov {\it et al.} (ZEUS Collaboration),
  %``Jet-radius dependence of inclusive-jet cross sections in deep inelastic
  %scattering at HERA,''
  Phys.\ Lett.\  B {\bf 649}, 12 (2007).
%  [arXiv:hep-ex/0701039].
  %%CITATION = PHLTA,B649,12;%%


%\cite{Glasman:2007sm}
\bibitem{Glasman:2007sm}
  C.~Glasman  (H1 Collaboration and ZEUS Collaboration),
  %``Precision measurements of alphas at HERA,''
  J.\ Phys.\ Conf.\ Ser.\  {\bf 110}, 022013 (2008).
  %[arXiv:0709.4426 [hep-ex]].
  %%CITATION = 00462,110,022013;%%


\end{thebibliography}
\end{document}